\documentclass[prc,twocolumn,epsfig,nofootinbib,floatfix]{revtex4}
\usepackage{graphics}
\usepackage{epsfig}
\usepackage{amsfonts}
\usepackage{amsmath}
\usepackage{bm}

\begin{document}
\title{
SELF-CONSISTENT SEPARABLE RPA FOR SKYRME FORCES: \\
GIANT RESONANCES IN AXIAL NUCLEI}
\author{V.O. Nesterenko$^1$, W. Kleinig$^{1,2}$, J. Kvasil$^3$,
P. Vesely$^3$, P.-G. Reinhard$^4$, and D.S. Dolci$^1$}
\affiliation{$^1$ Laboratory of Theoretical Physics,
Joint Institute for Nuclear Research, Dubna, Moscow region, 141980, Russia}
\email{nester@theor.jinr.ru}
\affiliation{$^2$ Technische Universit\"at Dresden, Inst. f\"ur Analysis,
   D-01062, Dresden, Germany}
\affiliation{$^3$ Institute of Particle and Nuclear Physics, Charles University, CZ-18000 Praha 8,
Czech Republic}
\affiliation{$^4$ Institut f\"ur Theoretische Physik II,
Universit\"at Erlangen, D-91058, Erlangen, Germany}

\date{\today}

\begin{abstract}
We formulate the self-consistent separable random-phase-approximation (SRPA)
method and specify it for Skyrme forces with pairing for the case of axially
symmetric deformed nuclei. The factorization of the residual interaction
allows to avoid diagonalization of high-rank RPA matrices, which dramatically
reduces the computational expense. This advantage is crucial for the systems
with a huge configuration space, first of all for deformed nuclei.
SRPA takes self-consistently into account the contributions of both
time-even and time-odd Skyrme terms as well as of the Coulomb force
and pairing. The method is implemented to description of isovector E1
and isoscalar E2 giant resonances in a representative set of deformed
nuclei: $^{154}$Sm, $^{238}$U, and $^{254}$No.
Four different Skyrme parameterizations (SkT6, SkM*, SLy6, and SkI3) are
employed to explore dependence of the strength distributions on some
basic characteristics of the Skyrme functional and nuclear matter.
In particular, we discuss the role of isoscalar and isovector
effective masses and their relation to time-odd contributions. High
sensitivity of the right flank of E1 resonance to different Skyrme forces
and the related artificial structure effects are analyzed.
\end{abstract}


\maketitle

\section{Introduction}
\label{sec:introduction}

Self-consistent mean-field models  with effective energy-density
functionals (Skyrme-Hartree-Fock, Gogny, relativistic) are established
as reliable tools for description of  nuclear structure and dynamics,
for a comprehensive review see \cite{Ben}. In particular, there is a
rising implementation of these models to dynamical features of nuclei
(see, e.g.
\cite{Stoitsov_PRC_03, Obertelli_PRC_05, Maruhn_PRC_05,Tere_PRC_05}),
which is caused, to a large extend, by exploiting the spectra and
reaction rates of exotic nuclei in astrophysics
\cite{Langanke_NPA_01,Stone_RPP_06}. However, applications of
self-consistent models to nuclear dynamics are still limited even in
the linear regime which is usually treated within the
random-phase-approximation (RPA). The calculations are plagued by
dealing with high-rank RPA matrices. This is especially painful for
deformed systems where lack of symmetry requires a huge
one-particle-one-hole $(1ph)$ configuration space.

The RPA problem becomes much simpler if the residual two-body
interaction is factorized (i.e. reduced to a separable form):
\begin{equation}
  \sum_{p_1h_1ph}\langle p_1h_1|V_{res}|ph \rangle \;
  a^+_{p_1} a^+_{h_1} a_p a_h
  \rightarrow
  \sum_{k,k'=1}^{K} \kappa_{k,k'}{\hat X}_k{\hat X}_{k'}
\label{1}
\end{equation}
where
${\hat X}_k=\sum_{ph}\langle p|{\hat X}_k|h \rangle a^+_p a_h$
is a hermitian $1ph$ operator and $\kappa_{kk'}$ is a matrix of
strength constants.  The factorization allows to reduce a high-rank
RPA matrix matrix to much smaller one with a rank 4K (where the
coefficient 4 is caused by the isospin and time-parity factors,
see discussion at the next Section).
The separable expansion can be formulated in a fully self-consistent
manner and provide a high accuracy with a small number (K $\sim 2 - 6$)
of the separable terms.

Several self-consistent schemes for separable expansions have been proposed
during the last decades \cite{BM,Row,LS,SS,Kubo,Ne_PRC,Vor,Sev}.
However, these schemes were not sufficiently general. Some of them are
limited to analytical or simple numerical estimates
\cite{Row,BM,LS,SS}, the others are not fully self-consistent
\cite{Ne_PRC}. Between them the approach \cite{Vor,Sev} for
Skyrme forces is quite promising. However, it still deals with RPA matrices
of rather high rank ($\sim 400$).

Recently, we have developed a general self-consistent separable RPA
(SRPA) approach applicable to arbitrary density- and current-dependent
functionals
\cite{Ne_PRA_98,Ne_AP_02,srpa_PRC_02,srpa_Miori,srpa_Prague,srpa_Houches,prep_05}.
The method was implemented to the Skyrme functional
\cite{Ben,Skyrme,Engel_75,Dob} for both spherical
\cite{srpa_PRC_02,srpa_Miori,srpa_Prague,srpa_Houches} and deformed
\cite{prep_05} nuclei.  In SRPA, the one-body operators ${\hat X}_k$
and associated strengths $\kappa_{k,k'}$ are unambiguously derived
from the given energy-density functional. There remain no further
adjustable parameters.  The success of the expansion depends on an
appropriate choice of the basic operators ${\hat X}_k$.  Experience
with spherical SRPA gives guidelines for an efficient choice such that
a few separable terms suffice to reproduce accurately the exact RPA
spectra \cite{srpa_PRC_02,srpa_Miori,srpa_Prague}.  The success
becomes possible due to the following factors:
i) an efficient self-consistent procedure \cite{Row,LS} based on sound
physical arguments,
ii) proper inclusion of all parts of the residual interaction,
    time-even as well as time-odd couplings,
iii) incorporation of the symmetries (translation, particle number,
     ...) leading to a correct description of the related
     zero-energy modes,
iv) building the separable
operators by such a way that they have maxima at different slices of
the nucleus and thus cover both surface and interior dynamics.
Furthermore, we note that SRPA is very general and can be applied to a
wide variety of finite Fermion systems. For example, SRPA was derived
for the Kohn-Sham functional \cite{KS,GL} and widely used for
description of linear dynamics of valence electrons in spherical and
deformed atomic clusters
\cite{Ne_PRA_98,Ne_EPJD_98,Ne_PRL_sciss,Ne_AP_02,Ne_EPJD_02,Ne_PRA_04}.

The enormous reduction of computational expense by SRPA is
particularly advantageous for deformed systems where 1ph configuration
space grows huge. SRPA becomes here a promising tool for large scale
studies. It is the aim of this paper to present a first implementation
of SRPA for axially deformed nuclei with pairing.
Like the full RPA, SRPA follows two strategies to compute
the dynamical response. It can be calculated via determination
of RPA eigenvalues and eigenstates or by a direct computation of the
multipole strength functions related to experimental cross
sections. We will discuss both ways.

As a first test, we will apply the method to description of
isovector E1 and isoscalar E2 giant resonance (GR) in the deformed
nuclei $^{154}$Sm, $^{238}$U, and $^{254}$No. These nuclei cover
a broad size range from rare-earth to super-heavy elements.
Four different Skyrme forces (SkT6 \cite{skt6}, SkM* \cite{skms},
SLy6 \cite{sly6}, and SkI3 \cite{ski3}) will be used to explore
dependence of the results on the actual parametrization. For our aims it is
important that these forces represent different values of
some relevant nuclear matter characteristics (isoscalar and isovector
effective masses and asymmetry energy). We will discuss
dependence of the collective strength on these characteristics,
scrutinize the impact of
time-odd coupling terms in the Skyrme residual interaction
and demonstrate important role of the Landau fragmentation.

The paper is organized as follows.  The general SRPA formalism is
presented in Sec. \ref{sec:srpa} and specified for Skyrme forces in
Sec. \ref{sec:skyrme}.  In Sec. \ref{sec:Q_choice} we discuss the
choice of the input operators.  Results of the calculations are
analyzed in Sec. \ref{sec:results}. The summary is done in
Sec. \ref{sec:summary}. Some important details of the method are given
in Appendices A-C. For more details, see documentation in
\cite{prep_05}.

\section{Basic SRPA equations}
\label{sec:srpa}

\subsection{The separable expansion}

In the most general form, the factorization of the residual interaction (\ref{1})
reads
\begin{widetext}
\begin{equation} \label{V_sep}
\hat{V}_{\rm res} \rightarrow \hat{V}_{\rm res}^{\rm sep} =
-
\frac{1}{2}\sum_{ss'}\sum_{k, k'=1}^{K} \{ \kappa_{sk,s'k'} {\hat X}_{sk}
{\hat X}_{s'k'}
+ \eta_{sk,s'k'} {\hat Y}_{sk} {\hat Y}_{s'k'} \}
\end{equation}
\end{widetext}
where the indices $s$ and $s'$ label neutrons and protons,
${\hat X}_{sk}$ are time-even hermitian one-body operators
and ${\hat Y}_{sk}$ are their time-odd counterparts,
$\kappa_{sk,s'k'}$ and  $\eta_{sk,s'k'}$ are the strength matrices.
Time reversal properties of the operators are
\begin{equation}
\begin{array}{rclclc}
T\hat{X}_{sk}T^{-1} &=& \gamma_T^X \hat{X}_{sk}&,& \gamma_T^X=+1 &,
\\
[3pt]
T\hat{Y}_{sk}T^{-1} &=& \gamma_T^Y \hat{Y}_{sk}&,&  \gamma_T^Y=-1 &,
\end{array}
\end{equation}
where $T$ is the operator of time reversal.  The expansion
(\ref{V_sep}) needs to take care of both classes of operators since
the relevant Skyrme functionals involve both time-even and time-odd
couplings, see \cite{Ben} and Appendix \ref{sec:dens_curr}.  Though
time-odd variables do not contribute to the {\it static} mean field
Hamiltonian of spin-saturated systems, they can play a role in
time-dependent perturbations and nuclear dynamics.  Altogether, the
presence of time-even and time-odd couplings in the functional
naturally leads to a formulation of the separable model in terms of
hermitian operators with a given time-parity.  These operators have
the useful property
\begin{equation}\label{eq:<[A,B]>=0}
  \langle  [\hat{A},\hat{B}] \rangle \sim (1-\gamma_T^A\gamma_T^B)
\end{equation}
which means that the average value of the commutator at the ground state
$| \rangle$ is not zero only for
operators of the opposite T-parities $(\gamma_T^A=-\gamma_T^B)$.  This
property will be widely used in the following.

\subsection{Linearized time-dependent mean field}

RPA is the limit of small-amplitude harmonic vibrations around the
ground state. The dynamics is formulated in general by a time-dependent
variation on the basis of a given energy functional
$E(J_{s}^{\alpha}({\vec r},t))$
\begin{equation}\label{eq:E(J)}
  \langle \Psi(t) |\hat H| \Psi(t) \rangle
  \longrightarrow
  E(J_{s}^{\alpha}({\vec r},t))
  =
  \int {\cal H} ({\vec r, t})d{\vec r} \; .
\end{equation}
We deal with a set of densities $J_{s}^{\beta}({\vec r})$
where $\beta$ denotes the type of density (spatial density,
kinetic density, current, spin density, spin-orbit density, ...)
and $s$ labels protons and neutrons. For reasons of compact
notation, we combine the density type $\beta$ with the space point
${\vec r}$ into one index $\alpha$ such that
\begin{equation}
 \alpha\equiv\left(\beta,{\vec r}\right)
 \quad,\quad
 \sum_\alpha ...
 =
 \sum_\beta \int d{\vec r} ...
\end{equation}
The densities are related to the corresponding one-body operators
$\hat{J}^{\alpha}_s$ (see the list in Appendix \ref{sec:dens_curr})
as
\begin{eqnarray}
  J_{s}^{\alpha}(t)
  &=&
  \langle \Psi(t) |\hat J_{s}^{\alpha}| \Psi(t) \rangle
\nonumber\\
  &=&
  \sum_{h \epsilon s}v_h^2
  \varphi^*_h(t)\hat J_s^{\alpha}\varphi_h^{\mbox{}}(t) \; .
\label{eq:dens_J}
\label{J}
\end{eqnarray}

Further, the state $|\Psi (t)\rangle$ is the underlying
Bardeen-Cooper-Schrieffer (BCS) mean-field
state composed from the single-particle states
$\varphi_h^{\mbox{}}({\vec r},t)$ and the corresponding pairing
occupation amplitudes $v_h$.
The time-evolution is determined by variation of the action
$\langle\Psi (t)| i \partial_t|\Psi(t)\rangle
  -\int d{\vec r} {\cal H} ({\vec r}, t)$.
Till now we keep the occupation amplitudes
$v_h$ fixed at their ground state values and consider the variation of
the single particle states. This yields the time-dependent mean-field
equations as
\begin{equation}
    i\frac {d}{dt}\varphi_h
  =
  \hat{h}\varphi_h
\label{eq:tdhf}
\end{equation}
with the mean field Hamiltonain $\hat{h}$ being a functional of the
local and instantaneous densities $J_{s}^{\alpha}({\vec r},t)$.
The freezing of the occupation amplitudes $v_h$ somewhat inhibits
application of SRPA for vibrational modes with a strong pairing
impact (e.g. for low-lying modes in neutron-rich light deformed
nuclei) \cite{nonfixed_v}. However, in the present study we are
interested in giant resonances where pairing dynamics plays a minor role.

In the linear regime of small amplitude oscillations, the
time-dependent state consists of the static ground state $|\rangle$
and a small time-dependent perturbation
\begin{equation}
  |\Psi(t)\rangle
  =
  | \rangle
  +
  |\delta\Psi(t)\rangle
\end{equation}
where both $|\Psi(t)\rangle $ and $| \rangle$ are BCS states.
Hence, all dynamical quantities can be decomposed as a sum
of the stationary ground state and small time-dependent parts:
\begin{subequations}
\label{eq:dens_J(t)}
\begin{eqnarray}
  J_{s}^{\alpha}({\vec r},t)
  &=&
  \tilde{J}_{s}^{\alpha}({\vec r})+
  \delta J_{s}^{\alpha}({\vec r},t)
  ,
\\
  \delta J_{s}^{\alpha}({\vec r},t)
  &=&
  \langle \Psi(t)|{\hat J}_{s}^{\alpha}|\Psi(t) \rangle -
  \langle {\hat J}_{s}^{\alpha}\rangle
  .
\label{eq:J(t)_psi}
\end{eqnarray}
\end{subequations}
Inserting (\ref{eq:dens_J(t)}) into (\ref{eq:E(J)}) and keeping the terms
up to the first order in $\delta J_{s}^{\alpha}({\vec r},t)$,
one obtains the single-particle Hamiltonian
\begin{equation}\label{full_h}
 \hat h (t)  = \hat{h}_0 + \hat{h}_{\rm res}(t)
\end{equation}
with the static mean-field part
\begin{eqnarray}\label{eq:h_0}
  \hat{h}_0
  =
  \sum_{\alpha s}
  \frac{\delta E} {\delta J_{s}^{\alpha}}\hat{J}_{s}^{\alpha}
\end{eqnarray}
and the time-dependent response
\begin{eqnarray}\label{eq:h_resp}
  \hat{h}_{\rm res}(t)
  &=&
  \sum_{\alpha's'}
  \frac{\delta {\hat h}}
    {\delta J_{s'}^{\alpha'}}
  \; \delta J_{s'}^{\alpha'}(t)
\\
  &=&
   \sum_{\alpha s \alpha's'}
  \frac{\delta^2 E}
  {\delta J_{s}^{\alpha}\delta J_{s'}^{\alpha'}}\Big|_{J=\bar{J}}
  {\hat J}_{s}^{\alpha}\,\delta J_{s'}^{\alpha'}(t)
\end{eqnarray}
related to the oscillations of the system.
Note that the second functional derivative is to be taken
at the ground state density as indicated by the index $J=\bar{J}$.
This holds for all second functional derivatives
and so we will skip this explicit index in the following.
For the brevity of notation, we will also skip
the explicit dependencies on space coordinates
and come back to these details in Sec. \ref{sec:skyrme} where
the residual interaction for the Skyrme functional is worked out.

The linearized equation of motion reads
\begin{equation}
  \left(i\frac {d}{dt}-\hat{h}_0\right)|\delta\Psi\rangle
  =
  \hat{h}_{\rm res}| \rangle.
\label{eq:lintdhf}
\end{equation}
Further steps require a more specific view of
$|\delta\Psi\rangle$. This will be done in the next subsections from
different points of view, macroscopic and microscopic.

\subsection{Macroscopic part of SRPA}
\label{sec:macro_SRPA}
\subsubsection{Scaling perturbed wave function}

It is convenient to obtain the perturbed mean-field state
$|\Psi(t)\rangle$ by the scaling transformation \cite{LS}
\begin{equation}\label{eq:scaling}
  |\Psi(t) \rangle_s
  =
  \prod_{k=1}^K
  \exp [-iq_{sk}(t)\hat{P}_{s k}]
  \exp [-ip_{sk}(t)\hat{Q}_{s k}]
  | \rangle_s
\end{equation}
where both $|\Psi(t)\! \rangle_s $ and $| \rangle_s$ are the
Slater determinants and ${\hat Q}_{sk}(\vec{r})$ and
${\hat P}_{sk}(\vec{r})$ are generalized coordinate (time-even) and
momentum (time-odd) hermitian one-body operators. These operators
fulfill the properties
\begin{subequations}
\label{eq:Q_P}
\begin{eqnarray}\label{eq:Q}
\hat{Q}_{sk} =\hat{Q}_{sk}^{\dagger},\quad \gamma_T^Q=1,  \\
\label{eq:P}
\hat{P}_{sk} = i[\hat{H},\hat{Q}_{sk}]=\hat{P}_{sk}^{\dagger},\quad
\gamma_T^P=-1
\end{eqnarray}
\end{subequations}
where $\hat{H}=\hat{h}_0 +\hat{V}_{\rm res}$ stands for the full
Hamiltonian embracing both the one-body mean-field Hamiltonian
and the two-body residual interaction. The
commutator in (\ref{eq:P}) is assumed to be mapped into the one-body
domain (see, e.g., the mapping into $\hat{h}_{\rm res}$ in
Eq. (\ref{eq:h_XY})).
If the functional includes only time-even densities, then
$\hat{V}_{\rm res}$ does not contribute to the commutator and so
$\hat{H}$ can be safely replaced by $\hat{h}_0$.

\subsubsection{Separable operators and strength constants}
\label{sec:sepopform}

The operators (\ref{eq:Q_P}) generate time-even and
time-odd real
collective deformations $q_{sk}(t)$ and $p_{sk}(t)$.  Using
(\ref{eq:scaling}) and assuming only small deformations, the
transition densities (\ref{eq:J(t)_psi}) read
\begin{eqnarray}\label{eq:trans_dens}
\delta J_{s}^{\alpha}(t)&  =&
i \sum_k
\{ q_{sk}(t)\langle [\hat{P}_{s k},\hat J_{s}^{\alpha}] \rangle
\\
&& \qquad\;
+ p_{s k}(t)
\langle [\hat{Q}_{s k},\hat J_{s}^{\alpha}] \rangle\}
\nonumber
\end{eqnarray}
where, following (\ref{eq:<[A,B]>=0}), time-even densities
contribute only to responses
$\langle |[\hat{P}_{s k},\hat J_{s}^{\alpha}]|\rangle$
while time-odd ones only to
$\langle |[\hat{Q}_{s k},\hat J_{s}^{\alpha}]|\rangle$.
Then the response Hamiltonian (\ref{eq:h_resp}) recasts as
\begin{equation}\label{eq:h_XY}
\hat{h}_{\rm res}(t)  = \sum_{s k}
\{ q_{sk}(t) \hat{X}_{s k} +
 p_{s k}(t) \hat{Y}_{s k}\}
\end{equation}
where the time dependence is concentrated in the amplitudes
$q_{sk}(t)$ and $p_{sk}(t)$ while all time-independent terms
are collected in the hermitian one-body operators
\begin{subequations}
\label{eq:X_Y}
\begin{equation}
\label{eq:X}
\hat{X}_{s k} = \sum_{s'}\hat{X}_{s k}^{s'}
= i\sum_{\alpha' \alpha s'}
\frac{\delta^2 E} {\delta J_{s'}^{\alpha'}
\delta J_{s}^{\alpha}}
\langle [\hat{P}_{s k} ,{\hat J}_{s}^{\alpha}] \rangle
{\hat J}_{s'}^{\alpha'} ,
\end{equation}
\begin{equation}
\label{eq:Y}
\hat{Y}_{s k} = \sum_{s'}\hat{Y}_{s k}^{s'}
= i\sum_{\alpha' \alpha s'}
\frac{\delta^2 E} {\delta J_{s'}^{\alpha'}
\delta J_{s}^{\alpha}}
\langle [\hat{Q}_{s k} ,{\hat J}_{s}^{\alpha}] \rangle
{\hat J}_{s'}^{\alpha'}
\end{equation}
\end{subequations}
with the properties
\begin{subequations}
\label{XY-properties}
\begin{eqnarray}\label{X-properties}
  \hat{X} = \hat{X}^{\dagger},  \quad
  &\gamma_T^X=+1,& \quad
  \hat{X}^* = \hat{X},
\\
\label{Y-properties}
  \hat{Y} = \hat{Y}^{\dagger}, \quad
  &\gamma_T^Y=-1,& \quad
  \hat{Y}^*=-\hat{Y} .
\end{eqnarray}
\end{subequations}

Obviously, ${\hat X}_{sk}$ and ${\hat Y}_{sk}$ are the reasonable
candidates for the time-even and time-odd operators in
the separable expansion (\ref{V_sep}).
The operator ${\hat X}_{sk}$ involves contributions only from
the time-even densities while the operator ${\hat Y}_{sk}$ only
from the time-odd ones. The upper
index $s'$ in the operators (\ref{eq:X_Y}) determines the
isospin (proton or neutron) subspace where these operators act.  This
is the domain of the involved density operators ${\hat J}_{s'}^{\alpha'}$.

To complete the construction of the separable expansion (\ref{V_sep}),
we should yet determine the matrices of the strength constants
$\kappa_{sk,s'k'}$ and  $\eta_{sk,s'k'}$. This can be done through
variations of the basic operators
\begin{widetext}
\begin{subequations}
\label{eq:XYvar}
\begin{eqnarray}\label{eq:Xvar}
  \delta {X}_{s k}(t)
  &=&
  \langle \Psi(t)|{\hat X}_{sk}|\Psi(t) \rangle
  - \langle {\hat X}_{sk} \rangle  =
  i \sum_{s'k'}
  q_{s'k'}(t)\langle [\hat{P}_{s'k'},\hat X_{sk}^{s'}] \rangle
  =  - \sum_{s'k'} q_{s'k'}(t) \kappa_{s'k',sk}^{-1} \; ,
\\
\label{eq:Yvar}
  \delta {Y}_{s k}(t)
  &=&
  \langle \Psi(t)|{\hat Y}_{sk}|\Psi(t) \rangle
  - \langle {\hat Y}_{sk} \rangle
  =  i \sum_{s'k'} p_{s'k'}(t)
  \langle [\hat{Q}_{s'k'},\hat Y_{sk}^{s'}] \rangle
  =  - \sum_{s'k'} p_{s'k'}(t) \eta_{s'k',sk}^{-1}
\end{eqnarray}
\end{subequations}
where
\begin{subequations}
\label{eq:kappa-eta}
\begin{eqnarray}
\label{eq:kappa}
  \kappa_{s'k',sk}^{-1 }
  =
  \kappa_{sk,s'k'}^{-1}
  &=&
  - i \langle [\hat{P}_{s'k'},{\hat X}_{sk}^{s'}] \rangle
  =  \sum_{\alpha \alpha'}
  \frac{\delta^2 E}{\delta J_{s'}^{\alpha'}\delta J_{s}^{\alpha}}
  \langle [\hat{P}_{s k},{\hat J}_{s}^{\alpha}] \rangle
  \langle [\hat{P}_{s' k'},{\hat J}_{s'}^{\alpha'}] \rangle \; ,
\\
\label{eq:eta}
  \eta_{s'k',sk}^{-1 }
  =
  \eta_{sk,s'k'}^{-1 }
  &=&
  -i
  \langle [\hat{Q}_{s'k'},{\hat Y}_{sk}^{s'}] \rangle
  =  \sum_{\alpha \alpha'}
  \frac{\delta^2 E} {\delta J_{s'}^{\alpha'}\delta J_{s}^{\alpha}}
  \langle [\hat{Q}_{s k},{\hat J}_{s}^{\alpha}] \rangle
  \langle [\hat{Q}_{s' k'},{\hat J}_{s'}^{\alpha'}] \rangle.
\end{eqnarray}
\end{subequations}
\end{widetext}
Eqs. (\ref{eq:kappa-eta})
represent the elements of the symmetric matrices which are inverse
to the matrices of the strength constants in (\ref{V_sep}).
Indeed, Eqs. (\ref{eq:XYvar}) can be recast to
\begin{subequations}
\begin{eqnarray}
- \sum_{sk} \kappa_{s'k',sk} \delta{X}_{s k}(t) &=& q_{s'k'}(t) \; ,
\\
-  \sum_{sk} \eta_{s'k',sk} \delta{Y}_{s k}(t) &=& p_{s'k'}(t) \; .
\end{eqnarray}
\end{subequations}
From that we read off the response Hamiltonian (\ref{eq:h_XY}) as
\begin{eqnarray}\label{eq:h_dXdY}
  \hat{h}_{\rm res}(t)
  &=&
  - \sum_{s'k'} \sum_{sk} \big\{
   \kappa_{s'k',sk} \delta\hat{X}_{sk}(t) \hat{X}_{s'k'}
\\
  && \qquad\quad\quad\quad
  + \eta_{s'k',sk} \delta\hat{Y}_{sk}(t) \hat{Y}_{s'k'}
  \big\} \; ,
\nonumber
\end{eqnarray}
which leads within  the collective space of the
generators $\{\hat{P}_{s k}\;,\;\hat{Q}_{s k}\}$
to the same eigenvalue problem  as the separable
Hamiltonian
\begin{equation}
{\hat H}={\hat h}_0 + {\hat V}^{\rm sep}_{\rm res}
\label{eq:H_sep}
\end{equation}
with ${\hat V}^{\rm sep}_{\rm res}$ given in Eq. (\ref{V_sep})
(see \cite{Row,srpa_PRC_02}).

In principle, we already have in our disposal the macroscopic SRPA
formalism for linear regime of the collective motion in terms of
real harmonic variables
\begin{subequations}
\label{eq:q_p}
\begin{eqnarray}
q_{sk}(t)=\bar{q}_{s k} \cos(\omega t)
=\frac{1}{2}\bar{q}_{s k}(e^{i\omega t}+e^{-i\omega t}) \; ,
\label{eq:q_harm}\\
p_{sk}(t)=\bar{p}_{s k} \sin(\omega t)=
\frac{1}{2i}\bar{p}_{s k}(e^{i\omega t}-e^{-i\omega t}) \; .
\label{eq:p_harm}
\end{eqnarray}
\end{subequations}
Indeed, Eqs. (\ref{eq:X_Y}) and (\ref{eq:kappa-eta})
deliver the one-body operators and strength matrices for the
separable expansion of the two-body interaction. By substituting
the response Hamiltonian (\ref{eq:h_dXdY}) and the perturbed wave function
(\ref{eq:scaling}) into time-dependent HF equation (\ref{eq:lintdhf})
one gets the eigenvalue problem. The number $K$ of the
collective variables (and thus of the separable terms)
is dictated by the accuracy we need in the description of collective modes.
For $K=1$, the method in fact is reduced to the sum rule approach with
one collective mode
\cite{Re_AP_92}.  For $K>1$, we have a system of K coupled oscillators and
the method becomes similar to so-called local RPA \cite{Re_PRA_90,Re_AP_92}
suitable for description of branching and gross-structure of collective modes.

\subsection{Microscopic part of SRPA}

The macroscopic SRPA as outlined in section \ref{sec:macro_SRPA}
serves here as a convenient tool to derive the optimal separable
expansion. But it cannot  describe the Landau fragmentation
of the collective strength. For this aim, we should build the
microscopic part of the model. In what follows, we will consider
the eigenvalue problem and the direct computation of the strength
function.

The perturbation $|\delta\Psi\rangle$ belongs
to the tangential space of the variations of a mean field state.
For the pure Slater states, they are all conceivable
one-particle-one-hole ($1ph$) excitations
\begin{equation}
  \hat{A}^\dagger_{ph}
  =
  a^{\dagger}_p a^{\mbox{}}_h \; , \quad
\hat{A}_{ph}
  =
  a^{\dagger}_h a^{\mbox{}}_p \; .
\label{eq:1ph}
\end{equation}
In the BCS case, the elementary modes are reduced to the
two-quasiparticle ($2qp$) excitations
$\hat{A}^\dagger_{ph}\longrightarrow
\hat{A}^\dagger_{ij}=\hat{\alpha}^\dagger_i\hat{\alpha}^\dagger_{\bar j}$
where $\hat{\alpha}^\dagger_j$ generate the BCS quasiparticle state
$j$ and $\bar j$ is its time reversal. Just this case is employed
in our actual calculations. Both $1ph$ and $2qp$ excitations
have much in common and result in the same microscopic SRPA
equations (for exception of pairing peculiarities outlined in Appendix
\ref{sec:pairing}). So, in what follows, we will not distinguish
these two cases. In particular, we will use for both excitations one and the
same notation $ph$.

\subsubsection{Eigenvalue problem}
\label{sec:eigen}

To formulate the eigenvalue problem,  we will
exploit the standard RPA technique for the separable
Hamiltonian (\ref{eq:H_sep}) where the separable operators and strength
constants are delivered by the macroscopic SRPA, see Eqs.
(\ref{eq:X_Y}) and (\ref{eq:kappa-eta}). Following \cite{Row},
the collective motion is represented in terms of
\begin{subequations}
\label{Psi_Thouless}
\begin{eqnarray}
  |\Psi(t) \rangle_{\nu}
  &=&
  \exp{\left(\hat{C}^{\dagger}_\nu e^{-i\omega_\nu t}
       -\hat{C}_\nu e^{+i\omega_\nu t}\right)}|\rangle
  \;,
\\
  \hat{C}^{\dagger}_{\nu}
  &=&
  \sum_s \sum_{ph\in s}
  \left(c^{\nu -}_{ph}\hat{A}^{\dagger}_{ph}
   - {c^{\nu +}_{ph}}_{\mbox{}}\hat{A}^{\mbox{}}_{ph}\right)
  \;,
\label{eq:geneigen}
\end{eqnarray}
\end{subequations}
where $\hat{C}^{\dagger}_\nu$ creates the one-phonon eigenmode $\nu$; the
operators $\hat{A}^{\dagger}_{ph}$ and  $\hat{A}^{\mbox{}}_{ph}$ are
defined in  (\ref{eq:1ph}); the perturbed $|\Psi(t) \rangle_{\nu}$
and ground $| \rangle$ states have the form of  Slater determinants.
We employ here the Thouless theorem \cite{Thouless} which establishes
connection between two arbitrary  Slater determinants.
The wave function (\ref{Psi_Thouless}) is a microscopic counterpart of
macroscopic scaling ansatz (\ref{eq:scaling}).
But here we aim a fully {\it microscopic}
description of excitations covering all the $1ph$ space while in the
previous sections we considered {\it macroscopic} flow as a benchmark
for forming the separable interaction.

The time-dependent response Hamiltonian $\hat{h}_{\rm res}(t)$
for the mode $\hat{C}^{\dagger}_\nu e^{-i\omega_\nu t}$
and the separable interaction (\ref{V_sep}) reads
\begin{eqnarray}
\label{eq:hres_sep}
  \hat{h}_{\rm res}^{\nu}
  &=&
  \sum_{sk}{\hat X}_{sk}
  \underbrace{
   \sum_{s'k'}\kappa_{sk,s'k'}
   \langle [{\hat X}_{s'k'},\hat{C}^{\dagger}_\nu] \rangle
  }_{
   \mbox{$\bar{q}_{sk}^{\nu}$}
  }
\nonumber\\
  &&
  +
  \sum_{sk}{\hat Y}_{sk}
  \underbrace{
   \sum_{s'k'}\eta_{sk,s'k'}
   \langle [{\hat Y}_{s'k'},\hat{C}^{\dagger}_{\nu}] \rangle
  }_{
   \mbox{-i$\bar{p}_{sk}^{\nu}$}
  }
  \;,
\end{eqnarray}
where
\begin{subequations}
\label{eq:barpqdef}
\begin{eqnarray}\label{eq:barqdef}
  \bar{q}_{sk}^{\nu}
  &=&
   \sum_{s'k'}\kappa_{sk,s'k'}^{\mbox{}}
   \\
   && \cdot\sum_{s"}\sum_{ph \in s"}
    \langle ph|\hat{X}_{s'k'}^{s"} \rangle
    (c^{\nu -}_{ph}\!+\!{c^{\nu +}_{ph}}_{\mbox{}}) ,
\nonumber
\\
\label{eq:barpdef}
  \bar{p}_{sk}^{\nu}
  &=& i
   \sum_{s'k'}\eta_{sk,s'k'}^{\mbox{}}
\\
   && \cdot\sum_{s"}\sum_{ph \in s"}
    \langle ph|\hat{Y}_{s'k'}^{s"} \rangle
    (c^{\nu -}_{ph}\!-\!{c^{\nu +}_{ph}}_{\mbox{}}) .
\nonumber
\end{eqnarray}
\end{subequations}
Note that here the values $\bar{q}_{sk}^{\nu}$ and
$\bar{p}_{sk}^{\nu}$ are new objects generated by the eigenmode
$\hat{C}^{\dagger}_\nu$.  They serve for notation of
$\hat{h}_{\rm res}$ and have no relation to the collective
generators $\bar{q}_{sk}$, $\bar{p}_{sk}$ used in the section
\ref{sec:macro_SRPA}.
Substituting the ansatz (\ref{Psi_Thouless}) into the linearized
time-dependent HF equation (\ref{eq:lintdhf}) and employing the form
(\ref{eq:hres_sep}) for the response field yields the expression
for $1ph$ expansion coefficients
\begin{eqnarray}
  c^{\nu \pm}_{ph \in s}
  =
  -\sum_{s'k'}
  \frac{\bar{q}_{s'k'}^{\nu}
        \langle ph|\hat{X}^s_{s'k'}\rangle
        \mp i
        \bar{p}_{s'k'}^{\nu}  \langle ph|\hat{Y}^s_{s'k'}\rangle}
  {2(\varepsilon_{ph}\pm\omega_{\nu})} \; ,
\label{eq:c_pm_qp}
\end{eqnarray}
where $\varepsilon_{ph}$ is the unperturbed energy of $1ph$ pair.
In the derivation above we used the operator properties
(\ref{XY-properties}). Then  the matrix elements
$\langle ph|\hat{X}_{s'k'}^{s"} \rangle$ and
$\langle ph|\hat{Y}_{s'k'}^{s"} \rangle$ are real and image,
respectively, while all the unknowns $c^{\nu \pm}_{ph}$,
$\bar{q}_{sk}^{\nu}$ and $\bar{p}_{sk}^{\nu}$ are real.

Inserting the result (\ref{eq:c_pm_qp}) into
Eqs. (\ref{eq:barpqdef}) yields finally
the set of SRPA equations for the values
$\bar{q}_{sk}^{\nu}$ and $\bar{p}_{sk}^{\nu}$
\begin{eqnarray}
\label{eq:X_c_qp4}
\sum_{\bar{s}\bar{k}}  &
\{
\bar{q}_{\bar{s}\bar{k}}^{\nu}
[F_{s'k',\bar{s}\bar{k}}^{(XX)}- \kappa_{\bar{s}\bar{k},s'k'}^{-1}]
+\bar{p}_{\bar{s}\bar{k}}^{\nu}
 F_{s'k',\bar{s}\bar{k}}^{(XY)}
\}=0 \; ,
\nonumber\\
\sum_{\bar{s}\bar{k}}  &
\{
\bar{q}_{\bar{s}\bar{k}}^{\nu}
F_{s'k',\bar{s}\bar{k}}^{(YX)}
+\bar{p}_{\bar{s}\bar{k}}^{\nu}
[F_{s'k',\bar{s}\bar{k}}^{(YY)} - \eta_{\bar{s}\bar{k},s'k'}^{-1}]
\}=0
\end{eqnarray}
where
\begin{widetext}
\begin{equation}\label{eq:F_AB}
 F_{s'k',\bar{s}\bar{k}}^{(AB)}  = \alpha_{AB}
 \sum_s \sum_{ph \in s}
\frac{1}{\varepsilon_{ph}^2-\omega_{\nu}^2}
\left\{ \langle ph|\hat{A}^s_{\bar{s}\bar{k}} \rangle^*
   \langle ph|\hat{B}^s_{s'k'} \rangle
(\varepsilon_{ph}+\omega_{\nu})
+ \langle ph|\hat{A}^s_{\bar{s}\bar{k}} \rangle
 \langle \hat{B}^s_{s'k'}|ph \rangle
(\varepsilon_{ph}-\omega_{\nu})
\right\}
\end{equation}
\end{widetext}
with $A,B \in X,Y$ and
\begin{equation}
\nonumber
\alpha_{AB}=\left(\begin{array}{l}
{1, \quad\mbox{for} \; A=B}\\
{-i, \; \mbox{for} \; A=Y, B=X}\\
{i, \quad \mbox{for} \; A=X, B=Y}
\end{array}\right) \; .
\end{equation}
The system of linear homogeneous equations (\ref{eq:X_c_qp4}) has
a non-trivial solution only if its determinant is zero. This yields
the dispersion equation to obtain the RPA eigenvalues
$\omega_{\nu}$.

\subsubsection{Strength function}
\label{sec:strength_func}

When exploring the response of the system to time-dependent external
fields, we are often interested in the total strength function rather
than in the particular RPA states. For example,
giant resonances in heavy nuclei are formed by thousands of RPA states
whose contributions in any case cannot be resolved experimentally. In
this case, it is more efficient to consider a direct computation of
the strength function which avoids the details and crucially
simplifies the calculations.

For an external electric field of multipolarity $E\lambda\mu$, we
define the strength function as
\begin{equation}\label{eq:strength_function}
  S_L(\lambda\mu , \omega)
  =
  \sum_{\nu}
  \omega_{\nu}^{L} M_{\lambda\mu \nu}^2 \zeta(\omega - \omega_{\nu})
\end{equation}
where
\begin{equation}
\zeta(\omega - \omega_{\nu}) = \frac{1}{2\pi}
  \frac{\Delta}{(\omega - \omega_{\nu})^2 + (\Delta/2)^2}              
\label{eq:lorfold}
\end{equation}
is the Lorentz weight with an averaging parameter $\Delta$ and
\begin{equation}\label{eq:tr_me}
  M_{\lambda\mu \nu}
  = \frac{1}{\sqrt{2}}\sum_s e^{eff}_s
  \sum_{ph \in {s}}
  \langle ph|f_{\lambda\mu} \rangle
(c^{\nu +}_{ph,s}+c^{\nu -}_{ph,s})
\end{equation}
is the matrix element of $E\lambda\mu$ transition from the ground state to the
RPA state $|\nu \rangle$. Further, $e^{eff}_s$ is an effective charge
to be specified later. The operator of the electric external field
in the long-wave approximation reads
\begin{equation}
\hat{f}_{\lambda\mu} = e
\frac{1}{1+\delta_{\mu ,0}}
r^{\lambda} (Y_{\lambda \mu} +  Y^{\dag}_{\lambda \mu})
\; .
\end{equation}
The (normalized) amplitudes $c^{\nu \pm}_{ph,s}$ follow from
(\ref{eq:c_pm_qp}). Unlike the standard definition of the
strength function with $\delta (\omega - \omega_{\nu})$, we
exploit here the Lorentz weight. It is convenient to simulate
various smoothing effects.

The strength function (\ref{eq:strength_function}) can be recast
to the form which does not need information on the particular
RPA states \cite{Malov}.
For this aim, we will use the Cauchy residue theorem. Namely, the
strength function is represented as a sum of the residues for the poles
$z=\pm \omega_{\nu}$. Since the sum of all the residues (covering all
the poles) is zero, the residues with $z=\pm \omega_{\nu}$ (whose
calculation is time consuming) can be replaced by the sum of residues
with $z=\omega \pm i(\Delta /2)$ and $z=\pm \varepsilon_{ph}$ whose
calculation is much less expensive. The explicit derivation is given in
\cite{prep_05}. The final expression reads
$$
  S_L (\lambda\mu ,\omega )
  =
  \Im\left[
   \frac{z^{L}\sum_{\beta \beta'}
            F_{\beta \beta'}(z) A_{\beta}(z) A_{\beta'}(z)}
        {\pi F(z)}
  \right]_{z=\omega\!+\!i\Delta /2}
$$
\begin{equation}
\label{eq:sf_1}
+\sum_s (e^{eff}_s)^2 \sum_{ph \in s}
\varepsilon_{ph}^{L}
\langle ph|f_{\lambda\mu} \rangle^2 \zeta(\omega -\varepsilon_{ph})
\end{equation}
where $\Im$ means the image part of the value inside
the brackets,  $F(z)$ is the determinant of the RPA matrix  (\ref{eq:X_c_qp4})
with $\omega_{\nu}$ replaced by the complex argument $z$,
$F_{\beta \beta'}(z)$ is the algebraic supplement of the determinant,
 and
\begin{subequations}
\begin{equation}
\label{eq:A_X_1}
A_{sk}^{(X)}(z) =
\sum_{s'} e^{eff}_{s'} \sum_{ph \in {s'}}
\frac{\varepsilon_{ph}
      \langle ph|X^{s'}_{sk} \rangle \langle ph|f_{\lambda\mu} \rangle}
      {\varepsilon_{ph}^2-z^2} \; ,
\end{equation}
\begin{equation}
\label{eq:A_Y_1}
A_{sk}^{(Y)}(z) =
i \sum_{s'} e^{eff}_{s'} \sum_{ph \in s'}
\frac{\omega_{\nu}
       \langle ph|Y^{s'}_{sk} \rangle \langle ph|f_{\lambda\mu} \rangle}
      {\varepsilon_{ph}^2-z^2} \; .
\end{equation}
\end{subequations}
For the sake of brevity, we introduced in (\ref{eq:sf_1})
the new index $ \beta =\{ skg \}$ where $g$=1
for time-even and 2 for time-odd quantities.
For example,
$A_{sk\; g=1} = A_{sk}^{(X)}$ and  $A_{sk\; g=2}= A_{sk}^{(Y)}$.

The first term in (\ref{eq:sf_1}) collects the contributions of the
residual interaction. It vanishes at $V_{\rm res}=0$. The second term
is the unperturbed (purely two-quasiparticle) strength function.

\subsection{Basic features of SRPA}
\label{sec:general}

Before proceeding to further specification of the method, it is worth to
comment some its essential points.
\\
$\bullet$
To determine the unknowns $c^{\nu\pm}_{ph,s}$ of a full (non-separable) RPA,
one requires diagonalization of the matrices of a high rank equal to the size
of the $1ph$ basis. The separable approximation allows to reformulate the
RPA problem in terms of a few unknowns $\bar{q}_{\bar{k}}$ and
$\bar{p}_{\bar{k}}$ and thus to reduce dramatically
the computational effort. As is seen from (\ref{eq:X_c_qp4}),
the rank of the SRPA matrix is equal to $4K$ where $K$ is the number of the
separable operators.
\\
$\bullet$ The number of the SRPA states $|\nu>$ is equal to the number
of the relevant $1ph$ configurations used in the calculations. In
heavy nuclei, this number can reach
$10^4$-$10^5$.  Every state $|\nu>$ is characterized
by the particular set of the values $\bar{q}^{\nu}_{sk}$ and
$\bar{p}^{\nu}_{sk}$.
\\
$\bullet$ Eqs. (\ref{eq:X_Y}) and (\ref{eq:kappa-eta}) relate the
basic SRPA values with the initial
functional and input operators $\hat{Q}_{sk}$. After choosing
$\hat{Q}_{sk}$, all other SRPA equations are straightforwardly
determined following the steps
\begin{eqnarray}
&&\hat{Q}_{sk} \; \Rightarrow \;
\langle [\hat{Q}_{sk},{\hat J}^{\alpha}_s]\rangle  \; \Rightarrow \;
\hat{Y}_{sk}, \; \eta_{sk,s'k'}^{-1} \; \Rightarrow \;
\hat{P}_{sk}
\nonumber\\
&& \qquad \Rightarrow \;
\langle [\hat{P}_{sk},{\hat J}^{\alpha}_s]\rangle  \; \Rightarrow \;
\hat{X}_{sk}, \; \kappa_{sk,s'k'}^{-1} .
\label{eq:Q-initial}
\end{eqnarray}
As is discussed in Sec. \ref{sec:Q_choice}, the proper choice of
$\hat{Q}_{sk}$ is crucial to achieve a good convergence of the
separable expansion (\ref{V_sep}) with a minimal number of separable
operators. SRPA itself does not provide a recipe to get $\hat{Q}_{sk}$
but these operators can be introduced following intuitive physical
arguments.
\\
$\bullet$ SRPA restores the conservation laws (e.g. translational
invariance) violated by the static mean field. If a symmetry mode
has a generator $\hat{P}_{\rm sym}$, then we  keep the conservation law
$[\hat{H},\hat{P}_{\rm sym}]=0$ by including
$\hat{P}_{\rm sym}$ into the set of the input generators $\hat{P}_{sk}$
together with its complement $\hat{Q}_{\rm sym}=i[\hat{H},\hat{P}_{\rm sym}]$.
\\
$\bullet$ The basic SRPA operators can be expressed via
the separable residual interaction (\ref{V_sep}) as
\begin{equation} \label{eq:XY_V_res}
\hat{X}_{sk} = -i [\hat{V}^{\rm sep}_{\rm res}, \hat{P}_{sk}]_{ph}, \qquad
\hat{Y}_{sk} = -i [\hat{V}^{\rm sep}_{\rm res}, \hat{Q}_{sk}]_{ph}
\end{equation}
where the index $ph$ means the $1ph$ part of the commutator. It is seen that
the time-odd operator $\hat{P}_{sk}$ retains the time-even part of
$V_{\rm res}^{\rm sep}$
to build $\hat{X}_{sk}$. Vice versa, the commutator with the time-even operator
$\hat{Q}_{sk}$ keeps the time-odd part of $V_{\rm res}^{\rm sep}$ to
build $\hat{Y}_{sk}$.

\section{SRPA with the Skyrme functional}
\label{sec:skyrme}
\subsection{Skyrme functional}

We use the Skyrme functional \cite{Skyrme} in the form
\cite{Engel_75,Re_AP_92,Dob}
\begin{eqnarray}
   {E}  &=&  \int d{\vec r} ( {\cal H}_{\rm kin}
 +{\cal H}_{\rm Sk}(\rho_s,\tau_s,
   \vec{\sigma}_s,\vec{j}_s,\vec{\Im}_s)
\\
  && \qquad  + {\cal H}_{\rm C}(\rho_p) + {\cal H}_{\rm pair}(\chi_s) ) \; ,
\nonumber
\end{eqnarray}
where
\begin{eqnarray}
   {\cal H}_{\rm kin} &= &  \frac{\hbar^2}{2m} \tau ,
 \label{Ekin}
 \\
 {\cal H}_{\rm C}
 &=& \frac{e^2}{2}
 \int  d{\vec r}_1 \rho_p(\vec{r})
              \frac{1}{|\vec{r}-\vec{r}_1|} \rho_p(\vec{r}_1)
\nonumber\\
     &-&
     \frac{3}{4} e^2(\frac{3}{\pi})^\frac{1}{3}
                          [ \rho_p(\vec{r})]^\frac{4}{3} ,
\label{Ecoul}
\\
  {\cal H}_{\rm pair}(\chi_s)
  &=&
  \frac{1}{2}\sum_s V_{{\rm pair},s}\chi^*_s\chi^{\mbox{}}_s ,
\label{pair_functional}
\end{eqnarray}
\begin{eqnarray}
   {\cal H}_{\rm Sk} &= &
                 \frac{b_0}{2}  \rho^2
                  -\frac{b'_0}{2} \sum_s \rho_s^2
\\
     & &
    -\frac{b_2}{2} \rho (\Delta \rho)
     +\frac{b'_2}{2} \sum_s \rho_s (\Delta \rho_s)
\nonumber \\
        & &
          + \frac{b_3}{3}  \rho^{\alpha +2}
          - \frac{b'_3}{3} \rho^\alpha   \sum_s\rho_s^2
            \nonumber \\
     & &
     +  b_1 (\rho \tau - \vec{j}^2)
     - b'_1 \sum_s (\rho_s \tau_s - \vec{j}_s^2)
\nonumber\\
     & &
     - b_4 \left( \rho (\vec{\nabla}\vec{{\Im}})
      + \vec{\sigma} \cdot (\vec{\nabla} \times \vec{j})\right)
\nonumber\\
     & &
     -b'_4 \sum_s \left( \rho_s(\vec{\nabla} \vec{\Im}_s)
              + \vec{\sigma}_s \cdot (\vec{\nabla} \times \vec{j}_s) \right)
\nonumber
 \label{Esky}
\end{eqnarray}
are kinetic, Coulomb, pairing and Skyrme terms respectively. The
densities and currents used in this functional are defined in the
Appendices \ref{sec:dens_curr} and \ref{sec:pairing}.  Densities
without the index $s$ involve
both neutrons and protons, e.g.  $\rho=\rho_p+\rho_n$. Parameters $b_i$
and $\alpha$ are fitted to describe ground state properties of atomic
nuclei, see e.g. \cite{Ben}.

For the sake of brevity, we omit here derivation
of the Skyrme mean field which can be found elsewhere, e.g. in
\cite{Re_AP_92,srpa_PRC_02,prep_05},
but present only the main values entering the SRPA equations.

\subsection{Second functional derivatives}

The crucial ingredients of the Skyrme SRPA
residual interaction are the second functional derivatives entering
expressions for the basic operators (\ref{eq:X_Y}) and
strength matrices (\ref{eq:kappa-eta}).  They read
\begin{eqnarray}\label{dE_drho_drho}
&&\frac{\delta^2 E}{\delta \rho_{s_1} ({\vec r}_1)\delta \rho_s ({\vec r})}
    = b_0 - b_0'\delta_{ss_1}  - (b_2 - b_2'\delta_{ss_1})\Delta_{{\vec r}_1}
\nonumber\\
  && \quad
  + b_3 \frac{(\alpha +2)(\alpha +1)}{3} \rho^{\alpha} ({\vec r})
\nonumber\\
  && \quad
  -b_3' \big[ \frac{\alpha (\alpha -1)}{3} \rho^{\alpha -2}({\vec r}) \sum_{s_2}\rho^2_{s_2}({\vec r})
\nonumber\\
  && \quad
          + \frac{2\alpha}{3} \rho^{\alpha -1}\big({\vec r}) (\rho_s ({\vec r})
   + \rho_{s_1}({\vec r})\big)
\nonumber\\
  && \quad
+ \delta_{ss_1}\frac{2}{3}\rho^{\alpha} ({\vec r}) \big]
\nonumber \\
 && \quad
  - \delta_{s s_1} \delta_{s p}\frac{1}{3}(\frac{2}{\pi})^{1/3}[\rho_p({\vec r})]^{-2/3}
     \} \;
    \delta ({\vec r}-{\vec r}_1)
\nonumber \\
&& \quad
  + \delta_{s s_1} \delta_{s p} \;\frac{e^2}{|{\vec r}-{\vec r_1}|} \; ,
\\
\label{dE_drho_dtau}
&&\frac{\delta^2 E}{\delta \tau_{s_1} ({\vec r}_1)\delta \rho_s ({\vec r})}
 = [ b_1 - b_1'\delta_{ss_1} ]\delta ({\vec r}-{\vec r}_1) \; ,
\\ \label{dE_drho_dJ}
&&\frac{\delta^2 E}{\delta {\vec \Im}_{s_1} ({\vec r}_1)\delta \rho_s ({\vec r})}
 = [ b_4 + b_4'\delta_{ss_1} ]{\vec \nabla}_{{\vec r}_1} \delta ({\vec r}-{\vec r}_1)
\; ,
\\
 \label{dE_dJ_drho}
&&\frac{\delta^2 E}{\delta \rho_{s_1} ({\vec r}_1)\delta {\vec \Im}_{s} ({\vec r})}
 = -[ b_4 + b_4'\delta_{ss_1} ]{\vec \nabla}_{{\vec r}_1} \delta ({\vec r}-{\vec r}_1)
\end{eqnarray}
for time-even densities  and
\begin{eqnarray}\label{dE_dj_dj}
&&\frac{\delta^2 E}{\delta{j}_{s_1,m} ({\vec r}_1)\delta{j}_{s,n}({\vec r})}
=
\\
 &&\qquad 2 [- b_1 + b_1' \delta_{s s_1}] \delta_{mn}
  \delta ({\vec r}_1-{\vec r}) ,
\nonumber
\\
\label{dE_dj_dsigma}
&&\frac{\delta^2 E}{\delta{\sigma}_{s_1,m}({\vec r}_1)\delta{j}_{s,n}({\vec r})}
=
\\
&&\qquad
 -[ b_4 + b_4' \delta_{s s_1}]\epsilon_{mnl}
 {\nabla}_{{\vec r}_1, \; l}\delta ({\vec r}_1-{\vec r})
\nonumber
\end{eqnarray}
for time-odd densities. The last two terms in (\ref{dE_drho_drho}) represent
the exchange and direct Coulomb contributions. Further, the indices $m, n$ in
(\ref{dE_dj_dj}) and (\ref{dE_dj_dsigma}) run over the three basis
spatial directions of the chosen representation (in
our case the cylindrical coordinate system) and $\epsilon_{mnl}$ is the totally
antisymmetric tensor associated with the vector product.

\subsection{Presentation via matrix elements}
\label{sec:rep_me}

The responses in (\ref{eq:X_Y}) and (\ref{eq:kappa-eta}) are expressed
in terms of the averaged commutators
\begin{equation}\label{eq:<[A,B]>}
  \langle [\hat{A},\hat{B}]\rangle  \quad \mbox{with} \quad
 \gamma_T^A= - \gamma_T^B
\end{equation}
where $|\rangle$ is the quasiparticle vacuum.  Calculation of these
values can be considerably simplified if to express them through the
matrix elements of the operators $\hat{A}$ and $\hat{B}$.

To be specific, we associate the operator $\hat A$ with a time-even
operator ${\hat Q}$ or a time-odd operator ${\hat P}$, which have real
or imaginary matrix elements, respectively. Then
\begin{subequations}
\begin{equation}\label{eq:AB_T_ev}
\langle [\hat{Q}_{sk},\hat{B}_s] \rangle =  4i  \sum_{ph \in s}
\langle  ph|\hat{Q}_{sk} \rangle  \Im \langle  ph|\hat{B}_s \rangle  \; ,
\end{equation}
\begin{equation}
\label{eq:AB_T_odd}
\langle [\hat{P}_{sk},\hat{B}_s] \rangle  = -4  \sum_{ph \in s}
\langle  ph|\hat{P}_{sk} \rangle  \Re \langle ph|\hat{B}_s \rangle  \; ,
\end{equation}
\end{subequations}
where both $p$ and $h$ run over all single-particle states and
$\Im $ and $\Re $ mean the imaginary and real
parts of the values to the right. The pairing factors are included
in the single-particle matrix elements.

Then elements of the inverse strength matrices are real and read
\begin{subequations}
\begin{equation}
\label{eq:kappa_me}
\kappa_{s'k',sk}^{-1 }
=-i \langle [\hat{P}_{s'k'},{\hat X}_{sk}^{s'}] \rangle
=4i \sum_{ph \in s'}
\langle  ph|\hat{P}_{s'k'} \rangle  \;
\Re\langle  ph|\hat{X}_{sk}^{s'} \rangle  \; ,
\end{equation}
\begin{equation}
\label{eq:eta_me}
\eta_{s'k',sk}^{-1 }
=-i \langle [\hat{Q}_{k'},{\hat Y}_{k} \rangle
=4 \sum_{ph \in s'}
\langle ph|\hat{Q}_{s'k'} \rangle \;
\Im\langle ph|\hat{Y}_{sk}^{s'}] \rangle \; .
\end{equation}
\end{subequations}
The responses entering $\hat X$ and  $\hat Y$ in (\ref{eq:X_Y})
are also real and read
\begin{subequations}
\begin{equation}
\label{eq:R_T_ev}
{\cal R}^{\alpha}_{X,sk} = i  \langle [\hat{P}_{sk},\hat{J}^{\alpha}_s] \rangle
\\
=
-4i \sum_{ph \in s}
\langle ph|\hat{P}_{sk} \rangle  \; \Re \langle ph|\hat{J}^{\alpha}_s \rangle
\; ,
\end{equation}
\begin{equation}
\label{eq:R_T_odd}
{\cal R}^{\alpha}_{Y,sk} = i  \langle [\hat{Q}_{sk},\hat{J}^{\alpha}_s] \rangle
\\
= -4  \sum_{ph \in s}
\langle ph|\hat{Q}_{sk} \rangle \; \Im \langle ph|\hat{J}^{\alpha}_s \rangle
\end{equation}
\end{subequations}
where $\langle ph|\hat{J}^{\alpha}_s|\rangle$ are transition densities.

We actually deal with axially symmetric systems. The modes
then can be sorted into channels with a given angular-momentum projection $\mu$
to the symmetry axis $z$. For even-even nuclei  $\mu$ takes integer values.
The explicit expressions for the responses in cylindrical coordinates (see
definition of these coordinates in the Appendix \ref{sec:cylindr}) then read
\begin{eqnarray}\label{vec_j_Y}
\vec{j}_{Y, sk}(\vec{r}) &=&
 i \langle [\hat{Q}_{sk},{\hat{\vec j}}_s]\rangle
\\
&=&
  (\vec{e}_{\rho}j_{Y, sk}^{\rho}(\rho,z)
+ \vec{e}_{z}j_{Y, sk}^{z}(\rho,z))\cos\mu\theta
\nonumber\\
&+&
 \vec{e}_{\theta}j_{Y, sk}^{\theta}(\rho,z)\sin\mu\theta \; ,
\nonumber\\
\label{vec_s_Y}
\vec{s}_{Y, sk}(\vec{r}) &=&
 i \langle[\hat{Q}_{sk},{\hat{\vec s}}_s]\rangle
\\
&=&
  (\vec{e}_{\rho}s_{Y, sk}^{\rho}(\rho,z)
+ \vec{e}_{z}s_{Y, sk}^{z}(\rho,z))\sin\mu\theta
\nonumber\\
&+&
  \vec{e}_{\theta}s_{Y, sk}^{\theta}(\rho,z)\cos\mu\theta \; ,
\nonumber\\
\label{rho_X}
\rho_{X, sk}(\vec{r}) &=& i \langle [\hat{P}_{sk},{\hat \rho}_s]\rangle
= \rho_{X, sk}(\rho,z)\cos\mu\theta \; ,
\\
\label{tau_X}
\tau_{X, sk}(\vec{r}) &=& i \langle [\hat{P}_{sk},{\hat \tau}_s]\rangle
= \tau_{X, sk}(\rho,z)\cos\mu\theta \; ,
\\
\label{vec_J_Y}
\vec{\Im}_{X, sk}(\vec{r})
&=& i \langle [\hat{P}_{sk},{\hat{ \vec\Im}}_s]\rangle
\\
&=&
  (\vec{e}_{\rho}\Im_{X, sk}^{\rho}(\rho,z)
+ \vec{e}_{z}\Im_{X, sk}^{z}(\rho,z))\cos\mu\theta
\nonumber\\
&+&
 \vec{e}_{\theta}\Im_{X, sk}^{\theta}(\rho,z)\sin\mu\theta
\nonumber
\end{eqnarray}
where $\{\rho, z\}$-depending response components are real.
All the dependence on the spatial coordinates follows from
the transition densities entering the responses.

The response components involving $\sin\mu\theta$ obviously vanish
for modes with $\mu =0$.

Explicit expressions in cylindrical coordinates
for strength matrices, responses, transition densities, matrix elements,
Coulomb contributions and other SRPA values can be found in
\cite{prep_05}. It is to be noted that the present calculations do not
yet include the Coulomb contribution to the residual interaction.
This introduces the uncertainty reaching  up to $\sim$0.4 MeV in an
average peak position \cite{sil_PRC_06}. Such
effect is safely below the precision of the present investigation.

\section{Choice of initial operators}
\label{sec:Q_choice}

As was mentioned in Sec. \ref{sec:general}, SRPA starts with the choice
of appropriate generating operators $\hat{Q}_{sk}$, see the sequence
of the model steps in (\ref{eq:Q-initial}).  The SRPA formalism itself
does not provide these operators. At the same time, their proper
choice is crucial to get good convergence of the separable expansion
(\ref{V_sep}) with a minimal number of separable terms.  The
choice should be simple and universal in the sense that it can be
applied equally well to different modes and excitation channels.

We propose a choice inspired by physical arguments.  The main idea is
that the generating operators should explore different spatial regions
of the nucleus, the surface as well as the interior.  The leading
scaling generator should have the form of the applied external field
in the long-wave approximation, which is most sensitive to the surface
of the system. Since nuclear collective motion dominates in the
surface region, already this generator should provide a good
description.  Next generators should be localized more in the interior
to describe an interplay of surface and volume vibrations.  For
$E\lambda$ giant resonances in spherical nuclei, we used a set of generators
with the radial dependencies in the form of power and Bessel functions.
\cite{srpa_PRC_02}. In the present study for deformed nuclei,
we will implement, for the sake of simplicity,
only generators with the power radial dependence
\begin{equation}
  \hat{Q}_{k}({\vec r})
  = r^{\lambda + 2(k-1)}
  (Y_{\lambda\mu} (\Omega )+ h.c.) .
\label{eq:scale_op}
\end{equation}
The separable operators $\hat X_k$ and $\hat Y_k$ with $k=1$ are
mainly localized at the nuclear surface while the operators with
$k>1$ allow to touch, at least partly,  the nuclear interior. This
simple set seems to be a good compromise for the first calculations
of the giant resonances.
Our analysis shows that already two first operators with $k=1,2$ suffice
for a spectral resolution of 2 MeV, as discussed below.
In the next studies we plan to enlarge the list of the input generators
so as to cover properly the nuclear interior and to take into account
the coupling of the modes with different $\lambda$, induced by
the deformation.

\section{Results and discussion}
\label{sec:results}

\subsection{Aims and details of calculations}

The main aim of the present calculations is to demonstrate ability
of SRPA to describe multipole giant resonances (GR) in deformed nuclei
and to explore dependence of the resonance strength distributions
on the nuclear matter properties and some terms of the Skyrme functional.
The isovector giant dipole resonance (GDR) and isoscalar giant quadrupole
resonance (GQR) in the axially deformed nuclei $^{154}$Sm, $^{238}$U,
and $^{254}$No will be considered. We will employ a representative
set of Skyrme forces, SkT6 \cite{skt6}, SkM* \cite{skms}, SLy6
\cite{sly6}, and SkI3 \cite{ski3}, with different
features (isoscalar and isovector effective masses, asymmetry energy,
...).  These forces are widely used for description of
ground state properties and dynamics of atomic nuclei \cite{Ben}
including deformed ones.  As is seen from Table 1, the forces
span a variety of nuclear matter properties while all correctly
reproducing experimental values of the quadrupole moments in
$^{154}$Sm and $^{238}$U.
\begin{table}
\caption{\label{tab:skyrme}
Nuclear matter and deformation properties for the Skyrme forces
under consideration. The table represents the
isoscalar effective mass $m_0^*/m$, symmetry energy $a_{\rm sym}$,
sum rule enhancement factor $\kappa$,  isovector effective
mass $m_1^*/m=1/(1+\kappa)$,
and quadrupole moments $Q_2$  in $^{154}$Sm, $^{238}$U, and $^{254}$No.
Experimental values  of the quadrupole moment in
$^{154}$Sm and $^{238}$U are 6.6 b and 11.1 b,
respectively \protect\cite{Goldhaber}.
}
\begin{ruledtabular}
\begin{tabular}{cccccccc}
Forces & $m_0^*/m$ & $a_{\rm sym}$ [MeV]&
$\kappa$ & $m_1^*/m$ & \multicolumn{3}{c}{$Q_2$ [b]}\\
  & &  &  &  & $^{154}$Sm & $^{238}$U  & $^{254}$No \\
 SkT6  & 1.00 & 30.0 & 0.001 & 1.00 & 6.8 & 11.1 &  13.7 \\
 SkM*  & 0.79 & 30.0 & 0.531 & 0.65 & 6.8 & 11.1 & 14.0 \\
 SLy6  & 0.69 & 32.0 & 0.250 & 0.80 & 6.8 & 11.0 & 13.7 \\
 SkI3  & 0.58 & 34.8 & 0.246 & 0.80 & 6.8 & 11.0 & 13.7 \\
\end{tabular}
\end{ruledtabular}
\end{table}

The calculations use a cylindrical coordinate-space grid with
the spacing of 0.7 fm. Pairing is treated at the BCS level and
is frozen in the dynamical calculations. The collective response
for the GDR ($\lambda=1$) and GQR ($\lambda=2$) is computed with two
input operators (\ref{eq:scale_op}) at k=1 and 2. Both GR
are calculated in terms of the energy-weighted (L=1)
strength function (\ref{eq:sf_1})
with the averaging parameter $\Delta$= 2 MeV (as most suitable for
the comparison with the experiment). The factorization of the
residual interaction and the
strength function technique  dramatically reduce the computational
effort.  For example, by using a PC with CPU Pentium 4 (3.0 GHz)
we need about 25 minutes for the complete calculations of the
GDR in $^{238}$U, including 1 minute for computation of the
strength function itself.

The isovector GDR and isoscalar GQR are calculated with the effective
charges $e_p^{eff}=N/A$, $e_n^{eff}=-Z/A$ and $e_p^{eff}=e_n^{eff}=1$,
respectively. The isoscalar dipole spurious mode is located at 2-3 MeV.
The deviation from the desirable  zero energy is caused
by several reasons. First, we have neglected in the present study
the contribution from the Coulomb residual interaction. The second
(more important) reason is that the nucleus is treated in a finite
coordinate box, which artificially binds the center of mass.
Larger numerical boxes could help here but at quickly increasing
computational cost. In any case, it suffices for our present purposes
that the center-of-mass mode lies at a low energy and thus is
safely separated from the GDR.

We use a large configuration space including the
single-particle spectrum from the bottom of the
potential well up to $\sim +16$ MeV.   This results in
7000-10000 dipole and 11000-17000 quadrupole two-quasiparticle
configurations in the energy interval 0 - 100 MeV.
The relevant energy-weighted sum rules are exhausted by $85-95 \%$.
Such a basis is certainly enough for the present aims.

\subsection{Discussion of results}

Results of the calculations are presented in Figs. 1-3.
The first two figures compare the calculated GDR and GQR
with the available photoabsorption \cite{nd_e1_exp,u_e1_exp}
and $(\alpha, \alpha')$ \cite{e2_exp} experimental data.
It is seen that all four Skyrme forces provide in
general an appropriate agreement with the experiment.
So, SRPA indeed can give a robust treatment of the multipole GR
and, what is important, does this with a minimal computational
effort.

\begin{figure}
\includegraphics[height=10.5cm,width=8cm]{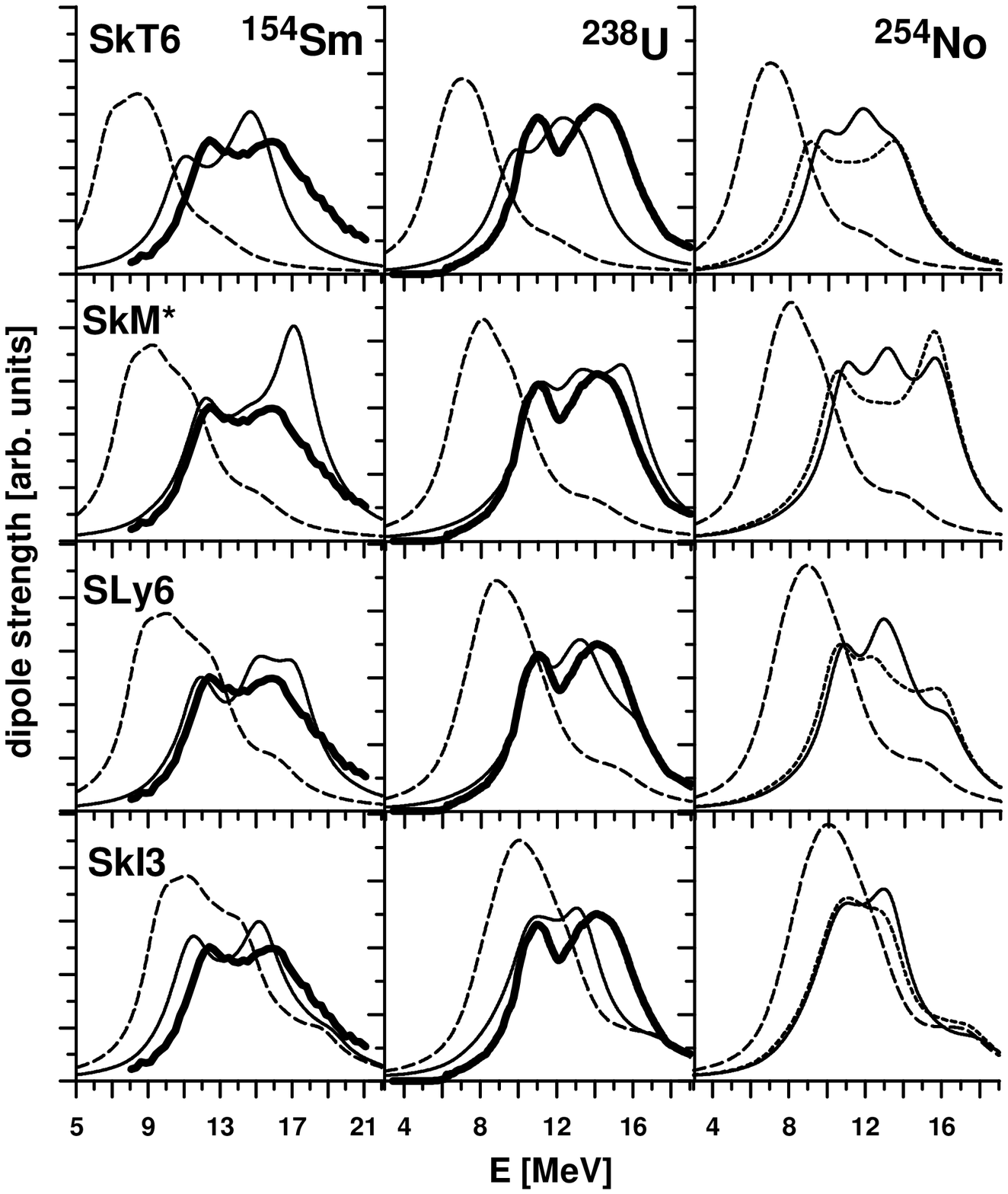}
\caption{\label{fig:E1}
The isovector GDR in $^{154}$Sm, $^{238}$U, and $^{254}$No
calculated with the Skyrme forces SkT6, SkM*, SLy6 and SkI3.
The plots depict: the collective strength calculated with two
(solid curve) and one (dotted curve for $^{254}$No) input operators,
 the unperturbed quasiparticle  strength (dash curve) and
the photoabsorption experimental data
\protect\cite{nd_e1_exp,u_e1_exp} (triangles).
}
\end{figure}
%
\begin{figure}
\includegraphics[height=10.5cm,width=8cm]{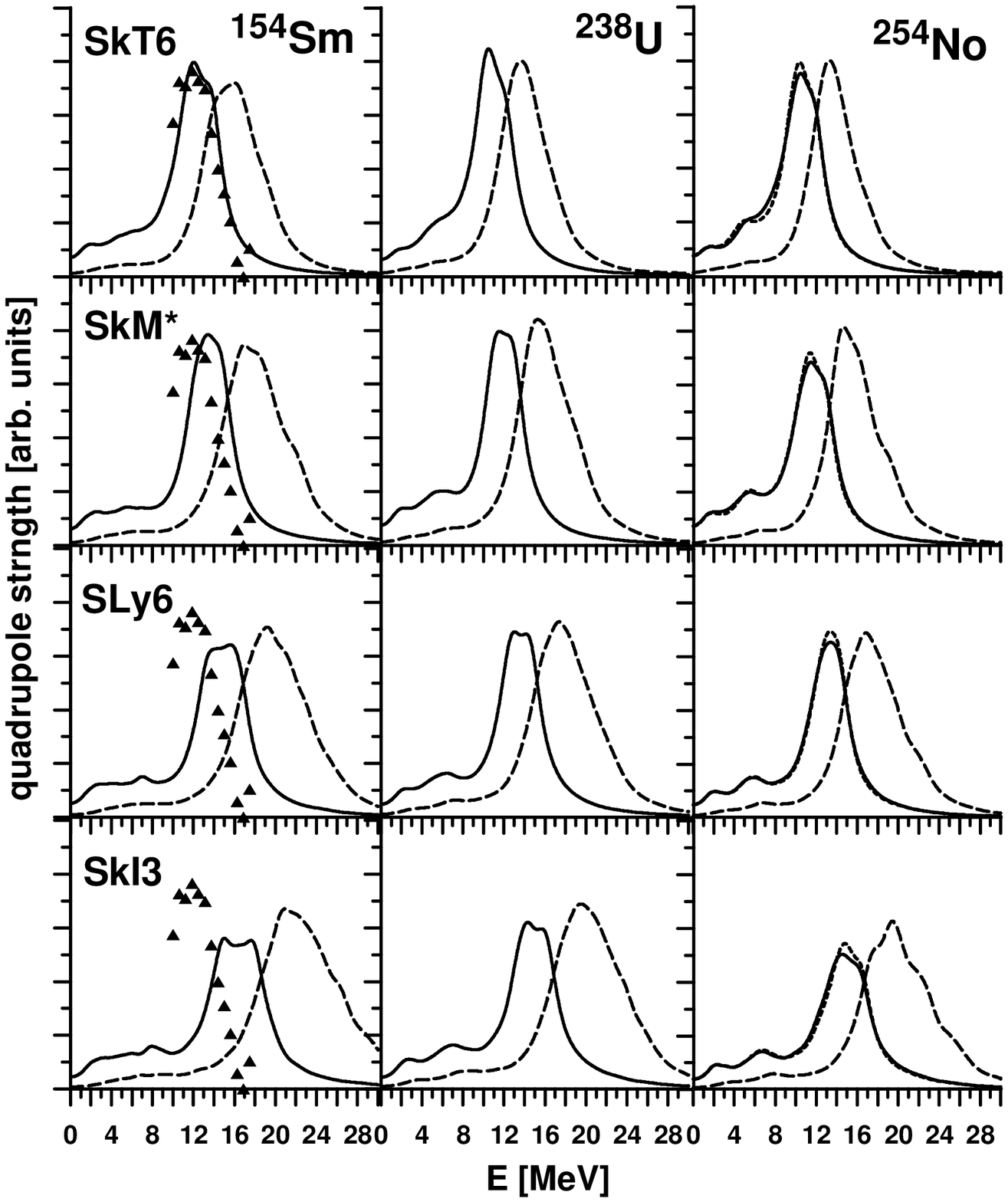}
\caption{\label{fig:E2}
The same as in Fig. 1 but for isoscalar GQR and $(\alpha, \alpha')$
experimental data \protect\cite{e2_exp}.
}
\end{figure}

To illustrate accuracy of the method, we give for $^{254}$No
the strength functions calculated with one (k=1) and two (k=1,2)
input operators. It is seen that both cases
are about equal for the GQR with its simple one-bump structure.
At the same time, the second operator considerably changes
gross structure of the more complicated GDR. We have checked that
inclusion of  more operators (k $\ge$ 3) does not result in
further significant modification of the GDR. So, the approximation
of two input operators seems to be reasonable, at least for the
present study at a resolution of 2 MeV.

The figures also show the unperturbed strengths, i.e. the mere
two-quasiparticle (2qp) distributions without the residual interaction.
Comparison of these strengths with the fully coupled collective
strengths (solid curves) displays the collective E1 and E2 shifts.
As a trivial
fact, we observe the proper right shift for the isovector GDR and
the left shift for the isoscalar GQR. What is more interesting,
the shifts for both resonances (including isovector DGR) depend on
the {\it isoscalar} effective mass $m_0^*/m$. To explain this, one
should remind that the smaller $m_0^*/m$, the more stretched the
single-particle spectrum \cite{Ben} (see also relevant examples in
\cite{nest_PRC_mom}). And indeed the E1 and E2 unperturbed strengths
in the figures exhibit a systematic shift to higher energy while
moving from SkT6 to SkI3 (for E1 the shift is essentially weaker than
for E2 \cite{eff_mass}). Simultaneously, we have the
corresponding evolution of the collective energy shifts.  Namely,
they decrease for GDR and increase for GQR (for exception
of GDR for SkT6).  All these trends lead to a remarkable net
result: strong variations of the unperturbed strengths and
collective shifts with $m_0^*/m$ considerably compensate each
other so as the final GDR and GQR energies become much less sensitive
to different Skyrme forces and approach the experimental values.

The remaining energy differences of order $\sim 1$ MeV  show some
other, sometimes known, trends for the GR.

First, we see a systematic (for exception of SkT6)
downshift of the calculated GDR  with increasing the symmetry energy
$a_{\rm sym}$. This, at first glance, surprising result complies
with  experience of the systematic studies in spherical nuclei
\cite{Rei_NPA_99}. The case of SkT6
looks as exception from this simple rule. It has the same asymmetry
energy as SkM$^*$ (see Table \ref{tab:skyrme}) but a lower GDR
resonance peak. This comes because the density dependence
$da_{\rm sym}/d\rho$ is abnormally low here (for reasons whose
discussion goes beyond the scope of the present paper).
Besides the trend with $a_{\rm sym}$, Fig. \ref{fig:E1} also
hints a connection of the GDR energy with the sum rule enhancement
factor $\kappa$ and the related value of the {\it isovector}
effective mass $m_1^*/m$. Namely, the smaller the
effective mass (larger the $\kappa$), the higher the GDR.
The trends and connections mentioned above should, however, be
considered with a bit of care. Indeed, the variation of $a_{\rm sym}$
between the different Skyrme forces in Table \ref{tab:skyrme}
is rather small and probably not enough to demonstrate a strong and
unambiguous trend. Besides, a complicated dependence of GDR on
{\it different} isovector factors can spoil and entangle the
concrete trends. In any case, analysis of the DGR trends
deserves more systematic study which is now in reach
with the efficient SRPA technique.

Instead,
the evolution of isoscalar GQR is more clear and systematic. We see a
steady upshift of the GQR peak from SkT6$^*$ to SkI3, which complies
with the known dependence on the isoscalar effective mass $m_0^*/m$
\cite{Bra_PReP_85}, namely, the lower the effective mass, the higher
the GQR. This trend has a simple explanation. As was discussed above,
the low $m_0^*/m$ results in stretching the single-particle spectrum and
thus in the upshift of the 2qp quadrupole strength. In the
GQR case, the opposite dependence of the collective shift is not enough
to compensate the strong 2qp upshift and hence we obtain the trend.
Note that in our calculations  SkT6 with $m_0^*/m=1$ yields the best
agreement for the GQR. This confirms to some extend the findings
\cite{Bra_PReP_85} for $^{208}$Pb that a good reproduction of the GQR
requires a large effective mass.

Shape and width of GR in deformed nuclei are mainly determined by
the deformation splitting and the spectral fragmentation due to
interference with energetically close $1ph$ states (Landau
fragmentation). Following Table \ref{tab:skyrme}, the different
Skyrme forces provide quite similar quadrupole moments. Hence
they result in close deformation splittings. Instead,
the Landau fragmentation depends sensitively on the spectral
pattern of a model, determined in a large extent by
isoscalar and isovector effective masses.
Our samples of forces contain sufficient variation of both ones.
Nonetheless, it turns out that the width and shape of the GQR
are practically the same for all the four Skyrme forces.

At the same time, the strongest variation is found for the GDR
whose width and gross-structure significantly depend on the force.
It is seen that SkM* gives an artificial third right peak
(especially in $^{154}$Sm) and thus an overestimation of the GDR
width. This effect weakens for SLy6 (leading to the best
description of GDR) and vanishes at all for SkI3 (resulting in
underestimation of the resonance width). The appearance of the
artificial right shoulder for SkM* and SLy6
was already noted for deformed rare-earth and actinide nuclei
\cite{Maruhn_PRC_05} and $^{208}$Pb \cite{srpa_PRC_02}.
For the particular Skyrme forces, this effect seems to be
universal. It takes place for GDR in heavy nuclei, independently on
their shape. As is shown in \cite{short}, the right shoulder is
provoked by an excessive collective shift
and further enforced by the presence in the
region of the 2qp bunch composed from the particular high-moment
 configurations ($\pi 1g_{9/2}$ and $\nu 1h_{11/2}$ for
$^{154}$Sm and $\pi 1h_{11/2}$ and $\nu 1i_{13/2}$ for $^{238}$U and
$^{254}$No) \cite{short}. These configurations
represent the intruder $l+1/2$ states entering the valence shell due to
the strong spin-orbital splitting. They form a dense 2qp bunch
which is easily excited and thus provides high sensitivity of the
right GDR flank. This feature can be useful for
additional selection of the parameters of Skyrme forces.

It would be very interesting to relate the above results
for the GDR width and profile with the effective masses.
Some possible correlations can immediately be noted.
For example,
the evolution of the right shoulder from SkM* to SkI3 seems
to correlate with a systematic decrease of $m_0^*/m$. Besides,
the largest right shoulder in SkM* can be related with
the smallest value of $m_1^*/m$ for this force.
This conclusion, however, requires still confirmation from
more extended studies with an even broader basis of forces.
At the same time, the results hint that
the properties of the isovector GDR probably depend not only
on the isovector $m_1^*/m$ but also on the isoscalar $m_0^*/m$.
For the first glance, this statement looks surprising.
However, we should take into account that $m^*_0/m$ influences
the mean level spacings $\bar \epsilon$ in the mean field
and hence can in principle affect the GDR.
The resonance should depend on the ground state properties
which in turn are related to $m^*_0/m$. It worth also noting that both
$m_0^*/m$ and $m_1^*/m$ are generated by one
and the same term of the Skyrme functional ($\sim b_1, b'_1$) and
so probably are not fully decoupled in the dynamics. Some non-trivial
relations between isoscalar and isovector parameters are already
discussed in literature, e.g. the relation
$\bar\epsilon m_0^*/m \approx \kappa m_1^*/m$
and its connection with $a_{sym}$  \cite{Satula}. And
we see here an interesting field for future investigations.
\begin{figure}
\includegraphics[height=10.5cm,width=7cm]{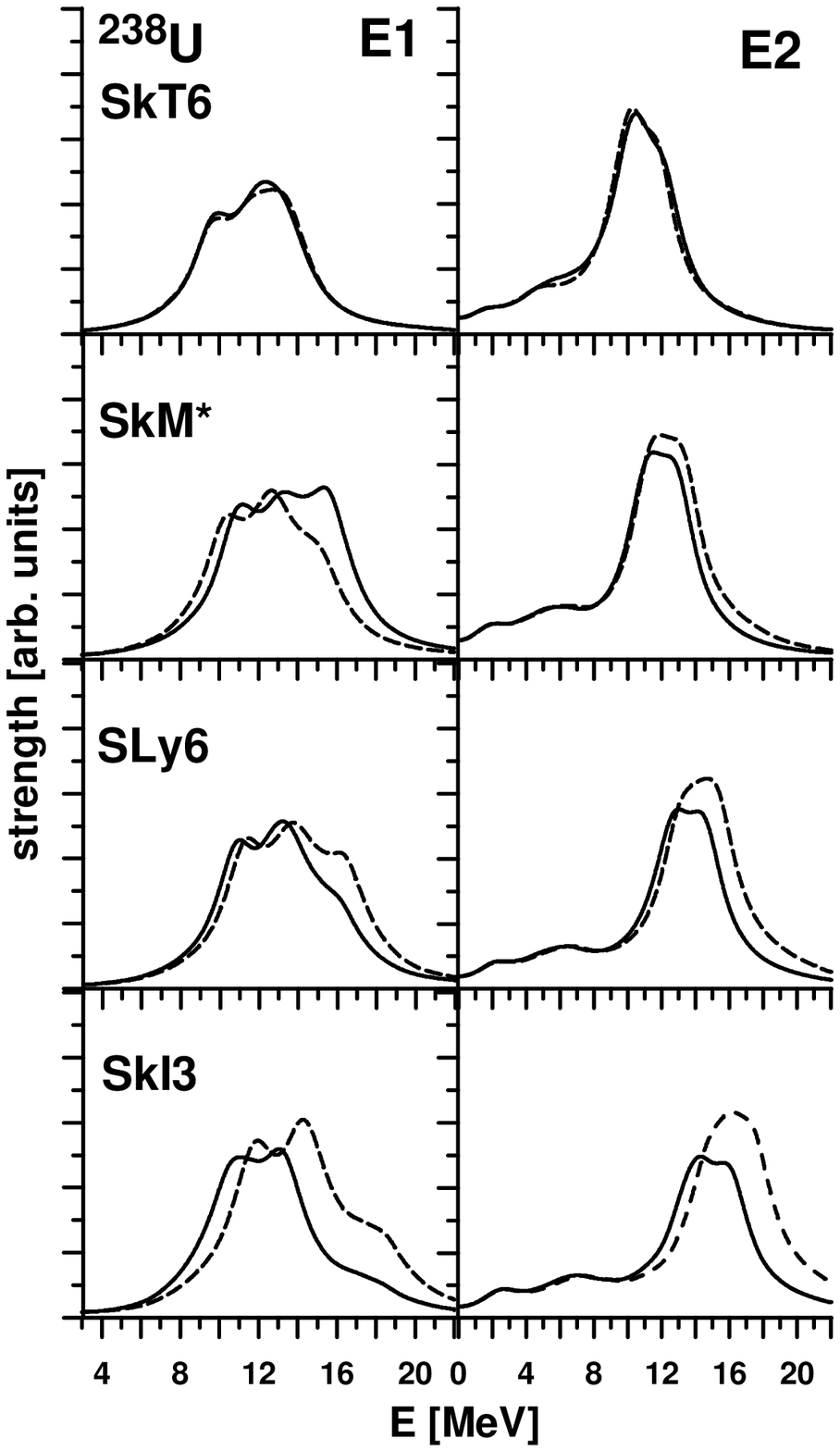}
\caption{\label{fig:todd}
The isovector GDR and isoscalar GQR in $^{238}$U calculated
with the Skyrme forces SkT6, SkM*,
SLy6 and SkI3. The strength is calculated with (solid curve)
and without (dash curve) contribution of the time-odd current.
}
\end{figure}

As a next step, let's consider contributions of the time-odd densities
to the GDR and GQR. These results are illustrated in
Fig. 3.  The calculations show that only the current-current contribution
(\ref{dE_dj_dj}) is essential while the contribution
(\ref{dE_dj_dsigma}) connected with the spin density is negligible.
So, in Fig. 3 we display only effect of the current density.
First of all, it is worth noting that the most significant changes appear
again at the right flank of the resonances. Probably, the
high-moment $l+1/2$-configurations mentioned above  play important
role not only for GDR but for GQR as well.

The impact of the time-odd current is different in GDR
and GQR. In the quadrupole resonance, we see the systematic
downshift and narrowing the strength. There is a clear correlation
with the value of the isoscalar effective mass: the lower $m_0^*/m$, the
stronger the time-odd effect. After inclusion of the time-odd coupling,
the GQR for different $m_0^*/m$ become much closer. So, the
time-odd contribution weakens the influence of $m_0^*/m$.

In the dipole resonance, the time-odd shift is not so systematic. We
observe no shift for SkT6, the upshift for SkM* and the downshifts for
SLy6 and SkI3. Again one may note some correlation with effective
masses.  The specific SkM* case can be connected with very small
$m_1^*/m$ for this force. Besides, one may note (for exception of
SkM*) increasing the time-odd impact with lowering $m_0^*/m$. In
principle, the correlations between the influence of time-odd terms
and effective masses in both GDR and GQR cases were
more or less expected
because the current density enters the term of the Skyrme functional
$\sim b_1, b'_1$ just responsible for generation of the effective
masses.  This explains why the SkT6 case with $m_0^*/m=m_1^*/m=1$ (no
effective mass effects) does not demonstrate any time-odd impact.

It is worth noting that the dominant contributions to the collective
response from the principle terms of the Skyrme functional have different
signs and thus, in a large extent, compensate each other (this can be easily
checked in the SRPA by estimation of the different contributions
to the inverse strength constants (\ref{eq:kappa-eta}).
As a result, the smaller contributions (time-odd,
spin-orbital and Coulomb) become important.

The present study involves three nuclei from different mass regions
and four Skyrme forces with various bulk properties. We have yet to
disentangle more carefully the separate influences of them. This
requires systematic variations of forces in a large set of test cases.
Such systematic studies can be now easily performed due to the
computationally efficient SRPA method.

\section{Conclusions}
\label{sec:summary}

A general procedure for the self-consistent factorization of the
residual nuclear interaction is proposed for arbitrary density- and
current-dependent functionals.  Following this procedure, the separable
RPA (SRPA) method is derived.  SRPA dramatically simplifies the
calculations while providing a reliable description of nuclear
excitations. The reduction of the computational effort is especially
useful for deformed nuclei. In the present paper, SRPA with Skyrme forces
is specified for description of collective dynamics in
axially-deformed nuclei.

For the first explorations, SRPA is applied for description of
isovector giant dipole resonances (GDR) and isoscalar giant quadrupole
resonances (GQR) in deformed nuclei from rare-earth ($^{154}$Sm),
actinide ($^{238}$U), and  superheavy ($^{254}$No) regions.  Four Skyrme
forces (SkT6, SkM*, SLy6, and SkI3) with essentially different bulk
properties are used. The calculations show that SRPA can successfully
describe multipole giant resonances.  A good agreement with
available experimental data is achieved, especially with  SLy6 for
GDR. We did not find any
peculiarities of the GR in superheavy nuclei. The behavior of
the resonances in all three mass regions looks quite similar.

We analyzed dependence of GDR and GQR  descriptions on various Skyrme
forces, in particular on the isoscalar and isovector effective masses
and the symmetry energy. The contribution of the time-odd couplings
was also explored. Some known trends for GDR with $a_{sym}$ and
GQR with $m_0^*/m$ were reproduced. Besides, the close relation
between the time-odd contribution and $m_0^*/m$ was demonstrated for GQR.
The calculations also hint some interesting (though not enough systematic)
trends for GDR. Altogether, the results point out correlations between
isoscalar and isovector masses and time-odd contributions. The
correlation seem to be natural since all the items originate from
one and the same term of the Skyrme functional.

The  time-odd and effective mass impacts manifest themselves mainly
at the right flanks of the strength distributions. The impacts are much
stronger and diverse for GDR. Besides, this resonance exhibits the strong
Landau fragmentation. The GDR gross structure considerably depends
on the applied Skyrme force and the related effective masses.
As a result, the GDR structure can serve as an additional
test for selection of the Skyrme parameters and related nuclear matter
values.

\vspace{0.2cm} {\bf Acknowledgments.}  This work is a part of the
research plan MSM 0021620834 supported by the Ministry of Education of
the Czech republic. It was partly funded by Czech grant agency (grant
No. 202/06/0363), grant agency of Charles University in Prague (grant
No. 222/2006/B-FYZ/MFF). We thank also the support from
DFG grant GZ:436 RUS 17/104/05 and
Heisenberg-Landau (Germany-BLTP JINR) grants for 2005 and 2006 years.
W.K. and P.-G.R. are grateful for the BMBF support under contracts
06 ER 808 and 06 DD 119.

\appendix 
\section{Density operators for Skyrme functional}
\label{sec:dens_curr}

In our study the Skyrme forces include
time-even (spatial, kinetic-energy, spin-orbit)
and time-odd (current, spin)  densities
associated with the hermitian operators
\begin{eqnarray*}
  \hat{\rho}_s(\vec{r})
  &=&
  \sum_{i=1}^{N_s}\delta(\vec{r}_i-\vec{r}) ,
\\
  \hat{\tau}_s(\vec{r})
  &=&
  \sum_{i=1}^{N_s}
  \overleftarrow{\nabla}\delta(\vec{r}_i - \vec{r})\vec{\nabla} ,
\\
  \hat{\vec{\Im}}_s(\vec{r})
  &=&
  \sum_{i=1}^{N_s}
  \delta(\vec{r_i} - \vec{r})\vec{\nabla}\!\times\!\hat{\vec{\sigma}} ,
\\
  \hat{\vec{j}}_s(\vec{r})
  &=&
  \frac{1}{2}\sum_{i=1}^{N_s}
  \left\{ \vec{\nabla},
  \delta(\vec{r}_i-\vec{r})
  \right\} ,
\\
  \hat{\vec{\sigma}}_s(\vec{r})
  &=&
  \sum_{i=1}^{N_s} \delta(\vec{r}_i-\vec{r})\hat{\vec{\sigma}} \; .
\end{eqnarray*}
where $\hat{\vec{\sigma}}$ is the Pauli matrix and $N_s$ is the number of protons
or neutrons.

The densities read as
\begin{equation}
\label{eq:dens}
  J_{s}^{\alpha} (\vec{r}) =
  \sum_{j \in s}v_j^2
  \varphi^*_j(\vec{r})\hat J_s^{\alpha}\varphi_j^{\mbox{}}(\vec{r})
\end{equation}
where $\hat J_s^{\alpha}$ is the density operator, $\varphi_j(\vec{r})$ is the
wave function of the single-particle state $j$, and $v_j^2$ is the pairing occupation
weight.

\section{Wave function in cylindrical coordinates}
\label{sec:cylindr}

Cylindrical coordinates ${\rho, z, \theta}$ are defined as
$$
x=\rho cos\vartheta , \;
y=\rho sin\vartheta , \;
z=z \; .
$$

Then the single-particle particle wave function and its time reversal
have the form of spinors
\begin{equation}
\varphi_j(\vec r) =
\left(\begin{array}{c} {R_{j}^{(+)}(\rho,z) e^{im_{j}^{(+)}\vartheta}} \\
{R_{j}^{(-)}(\rho,z) e^{im_{j}^{(-)}\vartheta}} \end{array}
 \right) ,
\end{equation}
\begin{equation}
\varphi_{\overline{j}}(\vec r) = \hat T \varphi_j(\vec r) =
\left(\begin{array}{c}
{-R_{j}^{(-)}(\rho,z) e^{-im_{j}^{(-)}\vartheta}} \\
{R_{j}^{(+)}(\rho,z) e^{-im_{j}^{(+)}\vartheta}} \end{array}
 \right)
\end{equation}
where $K_j$ is the projection of the complete single-particle
moment onto symmetry z-axis of the axial nucleus.

Expressions for differential operators in cylindrical coordinates are
elsewhere, see e.g. \cite{Korn}.

\section{Pairing contribution}
\label{sec:pairing}

The pairing is treated with Bardin-Coopper-Schriffer (BCS) method.
Then the SRPA values gain the pairing factors involving
coefficients $v_j$ and $u_j$ of the Bogoliubov transformation from
particles to quasiparticles.

The densities and currents in the Skyrme functional dominate in the
particle-hole channel. The pairing density falls into another channel
provided by the two-particle excitations. It involves a
different pairing weight, namely $u_j v_j$ rather
then the $v_j^2$ for the standard densities (\ref{eq:dens}).
Specifically the pairing density reads
\begin{equation}
  \chi_s
  =
  \sum_{j\in s}u_j v_j|\varphi_j|^2 \; .
\end{equation}

In the case of pairing, the hermitian one-body operators
with a given time-parity ($(\gamma_T^A=1,
\gamma_T^B=-1$) obtain in the $ph$-channel the form
\begin{eqnarray}\label{eq:Q_2qp}
{\hat A} &=&  2 \sum_{ij}
\langle ij|A \rangle u^{(+)}_{ij}
    (\hat{A}^\dagger_{ij}+\hat{A}_{ij}) \; ,
\\
{\hat B} &=&  2 \sum_{ij}
\langle ij|B| \rangle u^{(-)}_{ij}
(\hat{A}^\dagger_{ij}+\hat{A}_{ij}) \; ,
\label{eq:P_2qp}
\end{eqnarray}
where
\begin{equation}
  \hat{A}^\dagger_{ij} =
  \hat{\alpha}^{\dagger}_{i} \hat{\alpha}^{\dagger}_{\bar j} , \quad
\hat{A}_{ij}  =
\hat{\alpha}_{\bar j}\hat{\alpha}_{i}
\label{eq:2qp}
\end{equation}
are two-quasiparticle operators and
\begin{equation}\label{pairing_factors}
u^{(+)}_{ij}=u_i v_j + u_j v_i , \quad u^{(-)}_{ij}=u_i v_j - u_j v_i
\end{equation}
are the pairing factors. This is the case for time-even operators
$\hat Q_{sk}$ and $\hat X_{sk}$ and the time-odd operator $\hat Y_{sk}$.
The time-odd operator
\begin{eqnarray}
\hat P_{sk} &=& i[\hat H, \hat Q_{sk}]
=i \{ [\hat h_{0}, \hat Q_{sk}] + [\hat V_{\rm res}^{\rm sep}, \hat Q_{sk}] \}
\nonumber\\
&=& i [\hat h_{0}, \hat Q_{sk}] - \hat Y_{sk}
\end{eqnarray}
have a more complicated form because of the additional term
$i[\hat h_{0}, \hat Q_{sk}]$. Namely it reads
\begin{eqnarray}\label{eq:op_P_comm}
\hat{P}_{sk}
&=&
2 \sum_{ij \epsilon s}
\{
i2 \epsilon_{ij}  u^{(+)}_{ij} \langle ij|Q_{sk} \rangle
-  u^{(-)}_{ij}  \langle ij|Y_{sk}^s \rangle
\}
\nonumber\\
&& \cdot
(\hat{A}^\dagger_{ij}-\hat{A}_{ij})
\end{eqnarray}

The SRPA formalism in sections \ref{sec:srpa} and \ref{sec:skyrme}
is presented in a general form equally suitable for the cases with
and without pairing. In the case of pairing, the ground and
perturbed $1ph$ many-body wave functions are replaced by their BCS
counterparts, the summation indices $p$ and $h$ run the quasiparticle states,
and the involved values (densities, operators and matrix elements)
acquire the pairing factors given above.

The pairing is not important for giant resonances considered
in the present study. So, we freeze the pairing in the dynamics
and do not present here the explicit form of the pairing contribution
(pairing vibrations) to the residual interaction.
Some examples of this contribution can be found in \cite{prep_05}.


\begin{thebibliography}{99}
\bibitem{nonfixed_v} 
 We now work out the SRPA extension to solve this problem.
\bibitem{eff_mass}
The unperturbed E1 and E2 strengths are provided by transitions via one
and two quantum shells, respectively. Hence the stretching effect of the
effective mass $m_0^*/m$ is much stronger for E2 than for E1.
\end{thebibliography}
\end{document}